# LMS Moodle: Distance international education in cooperation of higher education institutions of different countries

*N. Kerimbayev, J.Kultan, S.Abdykarimova, A.Akramova*
*Al-Farabi Kazakh National University, Almaty, Kazakhstan*
*Economic Univesity v Bratislave, Bratislava, Slovak Republic*

**Abstract**
The development of international cooperation requires cooperation in the sphere of education. An enhanced sharing of experience in the sphere of practical teaching activities implies the increase of the quality of teaching process and of scientific cooperation. Sharing of experience in educational activities implies understanding among representatives of different nations anywhere in the world. It means that through LSM teaching the principle of social constructivism is realized, when participants together create a narrow culture of common objects and senses.

The article presents an example of practical application of electronic media to the process of a real lesson. The article describes the process of teaching students from different countries using the system of LMS MOODLE, beginning with preparing study materials, giving lectures by foreign lecturers, practical tasks and ending with passing an examination.

Training has included some full-time students from the Slovak Republic, the Republic of Dagestan and the Republic of Kazakhstan, and has been realized by applying the method of distance learning (LMS).

## 1. Introduction

One of forms of international cooperation is international exchange of students, exchange of teachers, teachers' trips to exchange experience. However, this kind of cooperation depends on finance problems, on the world economic crisis, etc. Disadvantages of the international cooperation in teaching include econimic problems, the necessity to cover vast distances, which needs a lot of time. These obstacles for cooperation can be overcome by using information technology in the form of webinars, video-conferences and systems of study material management.

These factors cause limited cooperation of teachers and students from different countries. Existing programs of academic mobility cover only a little part of teachers and students. How to realize and make accessible international cooperation, that is interaction of students and teachers from different countries?

We can solve this problem using IT, which provides with teaching students by foreign teachers and with students' participation in the educational process in different countries. A big advantage of applying IT to teaching-learning is the opportunity for large groups of students from different countries to communicate. Application of learning management systems /LMS Moodle/, video-systems and video-lectures or practical tasks using IT for communicating provides with creating a virtual environment close to a life situation (Kultan, 2009).

Now many different scientists discuss problems of developing and using. In the book "Building Virtual Communities of Practice for Distance Educators" (Bond M. A., Barbara B.. Lockee 2014) the authors rely on the development of a set of guidelines for creating a virtual community of distance teaching practice, which can easily be realized by specialists in the sphere of education.

The research of practical fundamentals of applying LMS platforms to teaching both young and older generations, creation of a stimulating contexts for future generations, relationship between distance learning and IT can be found in the works of Maria Sergeyevna Lyashenko (2014), Keengwe, Jared, and David Georgina (2012). Distance education is considered in the relationship between academic performance and students' technological adeptability. Many studies consider the impact of technology-assisted learning on academic performance among distance learners and their on-campus counterparts. Duvall, Cheryl King, and Robert G. Schwartz (2000) also study the relationship between academic performance and students' technological adeptability.

Since the appearance of new information and communication technologies (ICT), many have related to them as the new generation of distance education, and some have referred to their implementation in the academic community as challenging the very existence of campus-based universities. Today ICT significantly influences the development, education and socialization of children and youngsters. We can see this influence watching the rising generation use social networks, youngsters' sites not just for learning purposes. That is why there arises a need for developing and carrying out online activities for children and youngsters (Jonsson, Camilla, Simin Ghavifekr1 & Hazline Mahmood, 2015).

The role of distance education with the use of ICT is very important for lifelong learning. Today scientists explore the quest for a definition of lifelong learning, the recognition of the changing role of informal learning communities, proposing public policy considerations for those responsible for elementary education and the further development of information and communications technologies. (Kendall, Mike, 2005).

Distance education and information and communication technologies gave birth to the emergence of e-learning in the system of education and other spheres. 'Distance education' and 'e-learning' overlap in some cases, but they are not identical. "The lack of distinction between 'e-learning' and 'distance education' accounts for much of the misunderstanding of the ICT roles in higher education, and for the wide gap between the rhetoric in the literature describing the future sweeping effects of the ICT on educational environments and their actual implementation (Guri-Rosenblit S, 2005, Simin Ghavifekr1 & Hazline Mahmood, 2015)

The main software tool of electronic learning is lerning management system (LMS). LMS is a tool for organizing distance education.

In the article we attempt to familiarize readers with the process of teaching, in which the following higher education institutions took part: UIB (University of International Business, the Republic of Kazakhstan, Eurasian National University named after L.N.Gumilyev (Astana, Kazakhstan), University of Economics in Bratislava (Slovak Republic) and Dagestan State Institute of National Economy (DSINE, Makhachkala city). The organization of co-education of these higher education institutions demonstrates that distance, time difference and linguistic distinctions are not problems for establishing mutual cooperation, for sharing experience for the purpose of improving the educational process.

## 2. Methodology

The main objective of the information educational environment of LMS is raising the level and quality of methodological, didactic and information-related support of organizing an educational process for students and teachers.

Technically, e-Learning system is constructed on the basis of a wide range of software products.

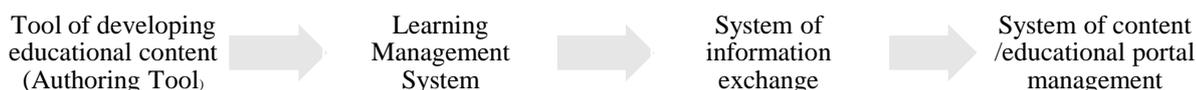

Tool of developing educational content (Authoring Tool) → Learning Management System → System of information exchange → System of content /educational portal management

The system of distance education is the basis of any e-Learning system.
The learning management system realizes the following functions
- Registration of listeners, personalization, access differentiation.
- Management of learning process, recording of learning outcomes and results of testing.
- Integration with mechanisms of synchronous and asynchronous communication.
- Integration with external information systems.

Using means of developing learning content, educational materials and tests are developed, which are included then in the database of the learning management system. Through this database listeners gain access to educational materials. The system of information exchange provides students, teachers, experts and other participants of the educational process with the possibility to exchange information both in real time (synchronously) and asynchronously. As a rule, web-interface of the learning management system is constructed on the basis of content management tools.

The educational content we have developed and use is one of the most popular and widely used Moodle platforms, it is interactive, it contains animation options and voiceover. For constructing the static content, some standard editing programs were used, such as Microsoft Word. The interactive content was created with a help of special software products.

The module of information exchange of the e-Learning system provided with the following functions (subject to the software chosen):
- Asynchronous communication, that is forums, bulletin boards, electronic mail;
- Synchronous communication, that is voice and computer chat, videoconferences, sharing of software products, virtual audience.

**3. Performance capabilities of educational materials in LMS MOODLE**

When teaching the course "Database Systems", the electronic course (Fig. 1) was developed in LMS MOODLE. The objective of this course is to familiarize students with basic concepts, to teach them to build up a database, to apply SQL instructions, to develop skills of solving problems of applying the data stored in a database.

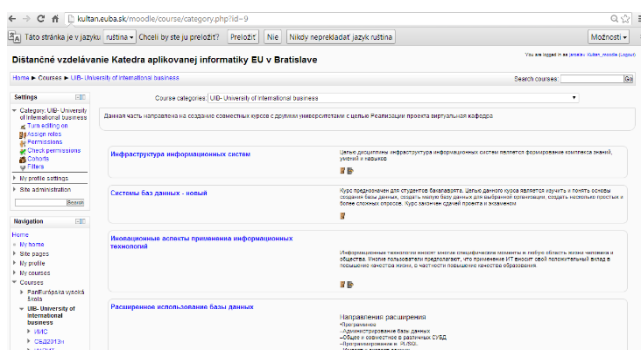

*Fig.* 1 *The courses listed in the website: kultan.euba.sk/moodle*

Thge electronic course was realized in the servers kultan.euba.sk/modle (or 193.87.20.2/moodle,) and http://virtualedu.kz. The data portion of the server contains several courses in the Kazakh, Slovak, Russian and English languages. They are applied in the teaching processes of the universities-partners. The educational material has been constructed by cooperation of teachers from UIB (Almaty) (http://uib.kz/), EUB (Bratislava) (http://euba.sk/), ENU (Astana) (http://enu.kz/), ISHK (Kazan) (http://isgz.ru/) and DSINE (Makhachkala) (http://dginh.ru/).

Besides the basic educational material presented in electronic form, the educational text of this course in Russian and Slovak was developed. Application of the electronic version of a textbook privides all the competent teachers with the possibility to make improvements in the textbook. Therefore, students can always have a current version. A paper version of the textbook provides students with possibility to use it if they do not have access to the Internet.

One of basic concepts of the system of distance learning Moodle is a course. Within the system a course is not only a means of organizing teaching process in the traditional understanding. A course can be just a communication area for interested people within particular subjects. A fulltime work at the system means setting up a computer account, which is compulsory. But depending on confugurations of each course, the access to it can be enlarged or limited.

The basic content of the course is divided into modules: zero module consisting of general for the whole course elements, and thematic modules.

Zero module usually contains course forums, course chats, general descriptions concerning the course in whole. Zero module of a typical educational course, for example, contains the forum

"Course News and Advertisements", the topics, which are automatically sent out to all the course participants. Only the teacher of a course can add a topic, but all the course participants can discuss it.

"Public Forum" and "Public chat" are intended for free communication of students and teachers. All the course participants can add and discuss topics; terminology and personality dictionaries.

A thematic module of a typical educational course, for example, can contain the following things:
- A brief description: dates of beginning and ending, subject, date of passing tests;;
- Notes of lectures and tests for self-control;
- Thematic test, trainings, final test (they are not shown in the picture).

The work at the forum has an activity-related module, which provides participants of a distance program with the possibility of asynchronous communication (Fig. 2). Within the system you can come across forums of different types:
- A standard forum, which consists of unlimited number of discussion topics and messages concerning these topics;
- A simple discussion, which consists of one topic and it is usually used for focusing discussion on one topic;
- A so called open topic, that is each of participants suggest one topic, participation in open topics is unlimited;
- The forum "question – answer". In this type of a forum only a teacher can create topics, a student will see the answers of other participants only after he/she answers the question proposed in the topic.

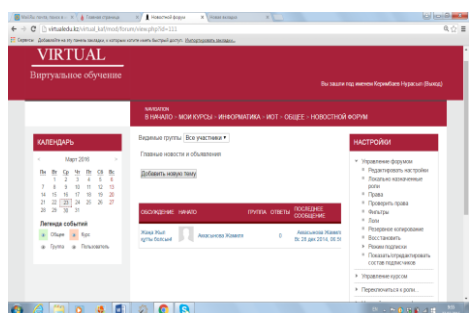

Fig. 2 Working with the forum

For each specific forum, a teacher or administrator can make the following configurations, which affect the work of users at it:

а) Access configurations: a standard public forum, that is a student can both create topics in the forum and give his/her answers in them; a news forum with the possibility of discussing topics, that is a student can only give his/her answers in topics but can not create them; a forum for distributing messages to uses, that is a student can neither create topics nor answer in them;

б) Possibility of subscription: all the participants are subscribed to a forum and can not be unsubscribed; participants can subscribe or cancel subscription to a forum; participants cannot subscribe to a forum;

в) Message assessment. For messages there can be an option of assessment, in this case the following additional configurations are possible:
- Only teachers can assess messages;
- Both teachers and students can assess messages;
- Students can see only their grades;
- Students can see all grades.

One of principle possibilities of a teacher (if you have all necessary rights) is editing of a course, that is adding, deleting, transferring resources, activity-related elements, and blocks.

Possibility is description of one typical function of a specific element of the system. For example, Course: Create, Task: Browsing, Forum: Subscription management. For each element of the system there are defined sufficiently many possibilities.

Permission is the value setting a certain possibility for a certain role. For example, to allow or to prohibit.

Context is some space of the Moodle system. For example, course, activity-related element, block.

The introduction page of electronic educational materials contains a general information of the given course, objectives of the course. At lectures students get tasks, which they have to send within a specific period of time. For those students who have some difficulties when solving problems there are some examples of problems solved. It is natural that the solutions without some novelty, which just follow a pattern, cannot be graded highly. When assessing tasks teachers should remember this. Main educational materials must be prepared before training sessions begin. Also one should remember that the given materials can be completed at any time during a course. In that way the drawbacks or errors, which are detected when performing particular tasks, can be corrected.

**Participants**

**Schedule for conducting training in Almaty**

The training schedule was developed in accordance with requirements and convenience of a foreign teacher. Lectures with the use of the LMS MOODLE system were scheduled.

When carrying out the training in Almaty city, most of the lectures were given within a month (the time of a foreign teacher's trip). For this aim, a subsidiary timetable for lectures and practical lessons was created. Additional lectures were scheduled for the time when students did not have other lessons, and then, after a foreign teacher left students had more free time. In addition, there is a possibility to use other teachers' lessons, which they can catch up later after a foreign teacher leaves. However, the latter variant is not optimal since then a lecture course is behind practical lessons.

Practical lessons were used for training in the remote access mode of a teacher (Fig.3). Students also worked through task passing mode, consultation mode, remote access mode and database management system.

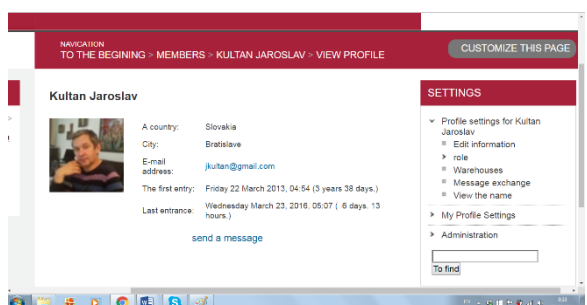

Fig. 3 Remote teacher

Further training was conducted using the method of distance learning at full-time department. Practical lessons were conducted once a week, according to the timetable. These lessons were aimed at explaining some tasks from the system of distance learning, checking up the tasks performed and realizing examples for explaining new tasks. At these consultations students can be checked up after solving short problems in the mode of direct connection with a chosen server.

The simplest form of connection is Skype with applying a camera (Fig. 3). All the students could be seen and heard, queries could be checked in the chat, and the results of these queries could be checked directly on the server of the database system. If needed we could apply the program Team viewer for viewing any student's computer screen.

The students were solving all the problems on the server Oracle, which was installed on the sever    kultan.euba.sk:8080/apex or MySQL installed on the site hostinger.ru (Рис.4).

Everybody decided to apply the server MySQL at the lessons. It was more convenient for the students since the language of communication was Russian. In the picture (Fig. 5), it is the beginning of the lesson. The teacher knew who presented at the lesson. In Fig. 6, you can see the content of the chosen server and created results of students' work. A teacher can see the results of each student's work in real time.

In such a way, lessons can last 50-100 minutes. You need not be online all the time, because students attend lessons being better prepared. As a rule, they know what they could not do, what problems they come across. But we think that you would better be online during the whole session /practical lesson/ for it makes students feel that they are in the interactive form of teaching-learning.

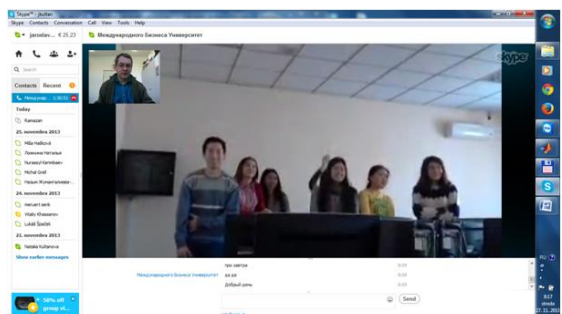 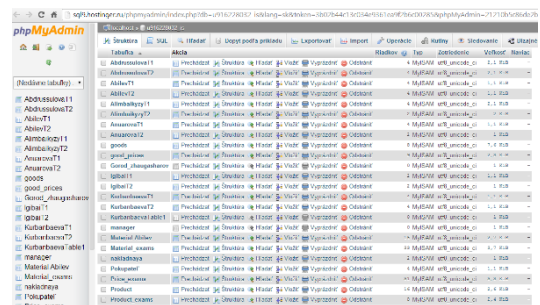

Fig. 5 Beginning of the lesson

Fig.6 Working field of the server Hostinger.ru

Since 2013 there has been a possibility to use the new development system (www.virtualedu.kz), which provides with a chance of organizing virtual (online) learning (Kerimbayev N.,2015, Kerimbayev N. et al., 2016).

The portal of "Virtual learning" is a developed virtual infrastructure including an electronic library, which provides with the possibility of participating in videoconferences in online mode. The specifics of this portal is that it provides with educational resources for trainees "at distance", gives additional possibilities of studying university subjects presented in the Virtual Center, original auctorial courses and study materials (Fig. 7).

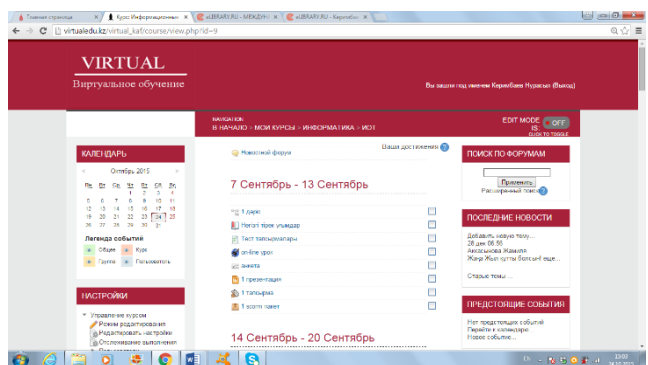

*Fig. 7. Resources of original auctorial courses*

The portal is represented by a text chat, using which one can communicate with other visitors during a conference. Participants exchange text messages using the chat, they have a chance to express their opinion and answer questions.

So called "on-line test" is carried out for controlling and self-controlling trainees'knowledge level. On-line tests help students determine not only their knowledge level but also the course of study for effective independent acquirement of a discipline.

As mentioned above, seminars are conducted during the whole semester. Students get used to these teaching methods step-by-step, and make better use of mutual communication facilities. The results of their work is better than those when using classical teaching methods. The electronic check-up and registration of a student's work make them participate in the lesson, not just be present.

**Schedule of training in Dagestan**

Training in this university is conducted in accordance with the schedule of Bachelor's full-time department. Besides, training is conducted under the developed system: students can see a teacher and a teaching material. The teacher simulteneously watches the activity in the classroom.

**Final examination**

At the end of each semester students' work is assessed. The examination and final rating include several components. A part of the grade is given for passed tasks and obtained results by a percentage (Fig. 8). A percentage is chosen as grade since it is difficult to determine the number of tasks at the beginning of the semester. The number of tasks can be increased or lessened based on the learning results gained by the students.

Fig. 8. The tasks passed and the results obtained during a semester

The examination consists of both theoretical and practical parts. The teacher on the site LMS (Fig. 9) familiarizes students with terms, time and methods of passing the examination. As students are not in the same room with the teacher, the examination tasks should be formulated so that they cannot be solved using the methods of CtrlC, CtrlV. Students get short answers. In the first part, it is necessary to work properly with systems of searching in formation. In the second part, which is aimed at analysis of the data gained, a student should demonstrate understanding of the problems. In the third part, he/she should apply and show that he/she can work with this instrument, that is being able to realize a certain part in his/her project.

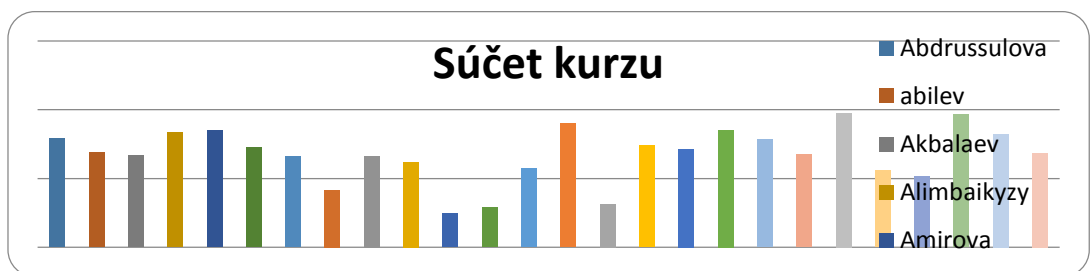

Fig. 9. Students' academic average

As to the theme, which must be developed, students get it at the very examination. There is one theme for each student. There is a tendency to give the same theme to two students with the aim to compare their results.

The practical part of the examination is conducted in the mode of direct connection with the server of the system chosen. Students can choose a server for better work. Students' tasks can be the same but should be formulated so that they do not suggest instructions of SQL. For example, "Write the instructions which show income from the sale of each group of products / services ".

students should apply these instructions to their database. Since each student has his/her own project, his/her own name and communication system, the probability of creating the same instruction is very small. Nevertheless, the teacher can check up each student's work using direct access methods to the computer desktop.

The examination is entirely supervised by the camera, and students know that the examination is recorded. It means that students are supervised all the time during an examination. That is the difference of such type of examination from the traditional method of examining.

The examination is conducted by distance method, the teacher can watch all the students' activities. In some cases, one can connect computers. Let us suppose that it is not necessary. A teacher can check up students' works on the real server.

5. **Conclusion**

The mentioned method of applying electronic systems to teaching provides with teachers' active participation in international system of teaching. Therefore, developed electronic courses and methods of teaching them can be used by other teachers, which increases international cooperation. Several universities can join to conduct joint courses in order to improve teaching and learning, and reduce costs.

Taking into account the results obtained the decision of shortening foreign teachers' presence at the university was made. At the beginning of an academic year, a fortnight is enough for the presence of a foreign teacher. Within this time, he will give introductory lectures and conduct introductory practical lessons. Besides, within this time all the passwords for access to LMS will be created. Also he can use the time of his colleague's lectures or practical lessons, for instance, on the course "Algorithms and their complexity". The courses will be taught in full-time regime and at practical lessons. We will take into account only time difference and timetables of UIB teachers in the event of connection failure.

Working experience in LMS MOODLE showed better results than when using traditional teaching methods. Besides, these new methods make both student and teacher be organized.

Electronic learning is not just a demonstration of our knowledge. It is an international need. Everything here depends on us. If we participate in international lecture-giving, those internships will improve our lectures' quality and, furthermore, will help spread our university's name.

**References**

Bond, M. A. & Lockee, B. B. (2014). *Building virtual communities of practice for distance educators*. New York: Springer

Duvall, Cheryl King, and Robert G. Schwartz. "Distance education: relationship between academic performance and technology-adept adult students." *Education and Information Technologies* 5.3 (2000): 177-187.

Jonsson, Camilla. "Are online communities for young people an issue for education researchers? A literature review of Swedish and international studies within the educational field." *Education and Information Technologies* 16.1 (2011): 55-69.

Ghavifekr, Simin, and Hazline Mahmood. "Factors affecting use of e-learning platform (SPeCTRUM) among University students in Malaysia." *Education and Information Technologies* (2015): 1-26.

Guri-Rosenblit S. 'Distance education'and 'e-learning': Not the same thing. *Higher education*. (2005) vol.49. №. 4. Pp. 467-493.

Keengwe, Jared, and David Georgina. "The digital course training workshop for online learning and teaching." *Education and Information Technologies* 17.4 (2012): 365-379.

Kendall, Mike. "Lifelong Learning Really Matters for Elementary Education in the 21st Century." *Education and Information Technologies* 10.3 (2005): 289-296.

Kerimbayev N. "Virtual learning: Possibilities and realization." *Education and Information Technologies* (2015): 1-13. Springer.

Kerimbayev N., Akramova A., and Suleimenova J. E-learning for ungraded schools of Kazakhstan: Experience, implementation, and innovation. *Education and Information Technologies*. (2016) 21.2. pp. 443-451. Springer.

Kultan, J. Zvýšenie kvality vzdelávania využitím spätnej väzby realizovanej pomocou LMS, Zborník príspevkov z medzinárodnej konferencie Inovačný proces v e-learningu 2009, Ekonomická univerzita v Bratislave, 24.03.2009 - 24.03. 2009,

Lyashenko, Maria Sergeyevna, and Natalja Hidarovna Frolova. "LMS projects: A platform for intergenerational e-learning collaboration." *Education and Information Technologies* 19.3 (2014): 495-513.